# Efficient Volumetric Absorption Solar Thermal Platforms Employing Thermally Stable - Solar Selective Nanofluids Engineered from Used Engine Oil

Nirmal Singh, Vikrant Khullar[*]

Mechanical Engineering Department, Thapar Institute of Engineering and Technology, Patiala-147004, Punjab, India

**ABSTRACT**
We report a low cost and scalable method to synthesize solar selective nanofluids from 'used engine oil'. The as-prepared nanofluids exhibit excellent long-term stability and photo-thermal conversion efficiency. Moreover, these were found to retain their stability and functional characteristics even after extended periods of high temperature (300°C) heating, ultra violet light exposure and thermal cyclic loading. Building upon it, we have been able to successfully engineer an efficient volumetric absorption solar thermal platform that employs the as-prepared nanofluids and achieves higher steady state temperatures (approximately 5% higher) relative to the conventional surface absorption based solar thermal system under the sun. The developed volumetric absorption solar thermal platform could prove to be significant step in the evolution of efficient solar thermal systems which could potentially be deployed for host of applications ranging from solar driven heating, air-conditioning, and desalination units to solar energy electricity generation systems.

**INTRODUCTION**
Nature has carefully engineered itself to utilize solar energy to run the entire life cycle on the planet earth. Mimicking natural processes, we are continuously striving to build efficient solar energy conversion platforms which could convert solar energy to more usable forms such as electricity (photovoltaic), chemical energy of the fuel (artificial photosynthesis) and heat (solar thermal). Amongst these, the technology of electricity generation via solar to heat conversion is currently the most efficient (~30%) and cost effective one. In comparison, the fossil-fuel based counterparts operate at efficiencies on the order of ~60%. One of the key reasons for lower efficiency is that we have not been able to efficiently engineer the solar (photon) to thermal energy (heat) conversion process [1].
With the advent of nanotechnology, plasmonic nanostructures (particularly metallic nanoparticles) have evolved as potential candidates for efficient photo-thermal conversion at their resonant frequencies owing to predominant non-radiative decay of the absorbed energy in the form of heat [2-4]. However, sunlight being broad spectrum necessitates plasmonic nanostructures (such as carbon-based nanostructures) that could respond to the wide wavelength range of incident sunlight [5-7].
Building upon the idea of utilizing nanostructures for solar to thermal energy conversion process; volumetrically absorbing solar thermal systems employing nanoparticle dispersions (nanofluids) have been devised by various researchers. Theoretically (and on laboratory scale), particularly at high solar concentration ratios (solar flux), the nanofluid-based volumetrically absorbing solar thermal systems have been shown to have higher thermal efficiencies, lower embodied energies and lower carbon footprints relative to their surface absorption based counterparts [8-11]. However, these promising novel systems have not been able to outperform the incumbent solar thermal platforms under the sun owing to inefficient receiver designs [12] and instability of nanofluids in real world service conditions - nanoparticles tend to agglomerate and hence settle down; this drastically affects the optical efficiency and hence the overall performance of these systems [13].

---

[*] Corresponding author
 Email address: vikrant.khullar@thapar.edu

Presently, a lot of efforts are underway to tailor solar selective, low cost, high temperature, and long term stable nanoparticle dispersions [14-28]. In this direction, we propose that 'used engine oil' (owing to the presence of carbon soot particles) could be employed to synthesize broad wavelength absorption nanoparticle dispersions (volumetrically absorbing solar selective heat transfer fluid).

Annually, approximately 24 million metric tons of 'used engine oil' is discharged into the environment without any recycling or treatment [29]. Therefore, forming one of the most hazardous wastes; having irreversible environmental and health implications. Putting this otherwise hazardous waste to harness solar energy could prove to be a sustainable option.

Pristine (or un-used) engine oil essentially consists of base oil (or blend of base oils) and an additive package to enhance its anti-oxidant, anti-wear, anti-foaming, and dispersancy characteristics. During its operation, the engine oil comes in contact with high temperature cylinder liners and washes away the carbon soot particles (left after combustion) from the cylinder circumference. Furthermore, a host of metallic particles (as a result of wearing action) enter the engine oil. The presence of dispersant molecules helps in dispersing the aforementioned foreign particles in the oil by forming an envelope around these particles. The polar part of the dispersant molecule attaches itself to the surface of the particle; whereas the oleophilic long chain hydrocarbon part helps in mobility of the particle. This ensures that the soot particles do not interact with each other and hence prevents agglomeration of the soot particles. The blowby gases also enter the crankcase which may tend to oxidize the engine oil; and here comes the role of anti oxidant which interupts the oxidation mechanism by reacting with the reaction intermediates [30-34].

After the end of the service life of the engine oil, in addition to resin, sludge etc.; it consists of large number of nano-sized soot particles which have undergone sort of 'functionalization' (as dispersant molecules envelope these soot particles). It is the essentially the presence of these 'functionalized carbon soot particles' in the used engine oil that qualifies it to be used as a precursor for synthesizing heat transfer fluid for direct absorption of solar energy.

Rigorous testing (which simulate real world service conditions) of the as-prepared nanofluids reveals that these have remarkable photo-thermal conversion efficiency, favorable thermo-physical properties (thermal conductivity and viscosity), high temperature and long term stability, and can withstand thermal cyclic loads without any significant loss of optical and thermo-physical properties. Building on it; a hybrid volumetric receiver employing the as-prepared nanoparticle dispersions has been carefully designed to give higher steady state temperatures (and hence higher thermal efficiency) relative to the conventional surface absorption based receiver under real world outdoor conditions.

In essence, the present work is a significant step in the evolution of solar thermal platforms; wherein we have been able to design a unique volumetric absorption based receiver that employs thermally stable and solar selective nanoparticle dispersions engineered from 'used engine oil' and has higher efficiency relative to its surface absorption based counterpart under the sun.

## RESULTS AND DISCUSSION
### Nanofluid synthesis philosophy and elemental-morphological analysis
In the present work, small fractions of 'used engine oil' (after undergoing filtration process) have been mechanically mixed with compatible non polar base oil (paraffin oil light). The mixture was then ultra-sonicated to get the required nanoparticle dispersions (see Fig. 1).

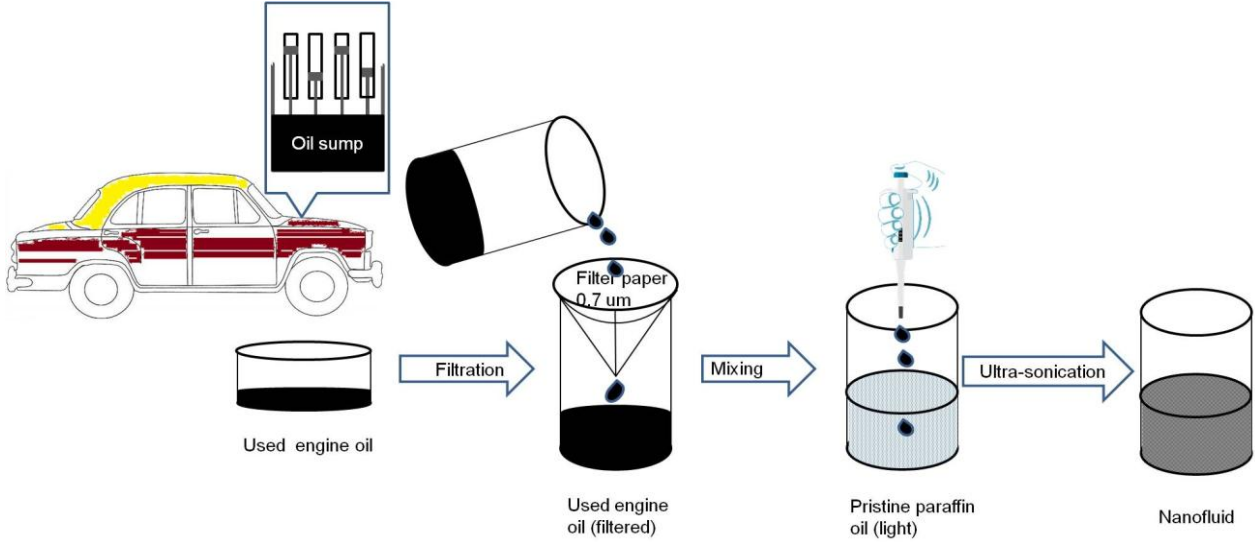

Fig. 1 Schematic showing the steps involved in nanoparticle dispersion synthesis.

Figure 2(a) shows the picture of as-prepared nanofluids of different concentrations (1.25, 2.5, 5, 10 and 20 ml/L) of 'used engine oil' in pristine paraffin oil. As shown in Fig. 2(b), through EDS, the percentage of carbon soot particles in the as-prepared nanoparticle dispersion has been found to be on the order of ~85% by weight (other notable elements being O, Al, Ca, and Fe). TEM images show that the soot particles are present in the form of nano-clusters of irregular curvilinear geometry [see Fig. 2(c)]. Hydrodynamic particle size has been measured through DLS; particle size varies in the range of 15 nm to 68 nm, average particle size being 38 nm [see Fig. 2(d)].

**Photo-thermal conversion efficiency**
As a first step, optical signatures of various nanofluid concentrations have been measured in the UV-VIS-NIR region (300 nm - 1100 nm). Figure 3(a) clearly shows that pure paraffin oil transmits nearly all the incident radiation whereas 20 ml/L nanofluid absorbs nearly in the entire wavelength band; thus giving us a fair idea about the absorption capability of the as-prepared nanofluids. This data was then employed to calculate the solar absorption fraction for different nanofluid concentrations as a function fluid layer thickness as

$$\text{Solar absorption fraction} = \frac{\int_{300nm}^{1100nm} S_\lambda \left[1 - \exp(-K_{a\lambda} y)\right] d\lambda}{\int_{300nm}^{1100nm} S_\lambda d\lambda}, \qquad (1)$$

where $S_\lambda$ is the spectral solar irradiance (AM 1.5 spectra), $K_{a\lambda}$ is the spectral absorption coefficient, and $y$ is the fluid layer thickness. Solar absorption fraction essentially gives the fraction of the incident solar energy that could be absorbed by a given thickness of the fluid layer. Clearly, solar absorption capability increases rapidly with increase in concentration. Moreover, to achieve the desired value of solar absorption fraction, we could either increase the concentration or increase the physical thickness of the nanofluid layer - in effect increasing the optical depth.

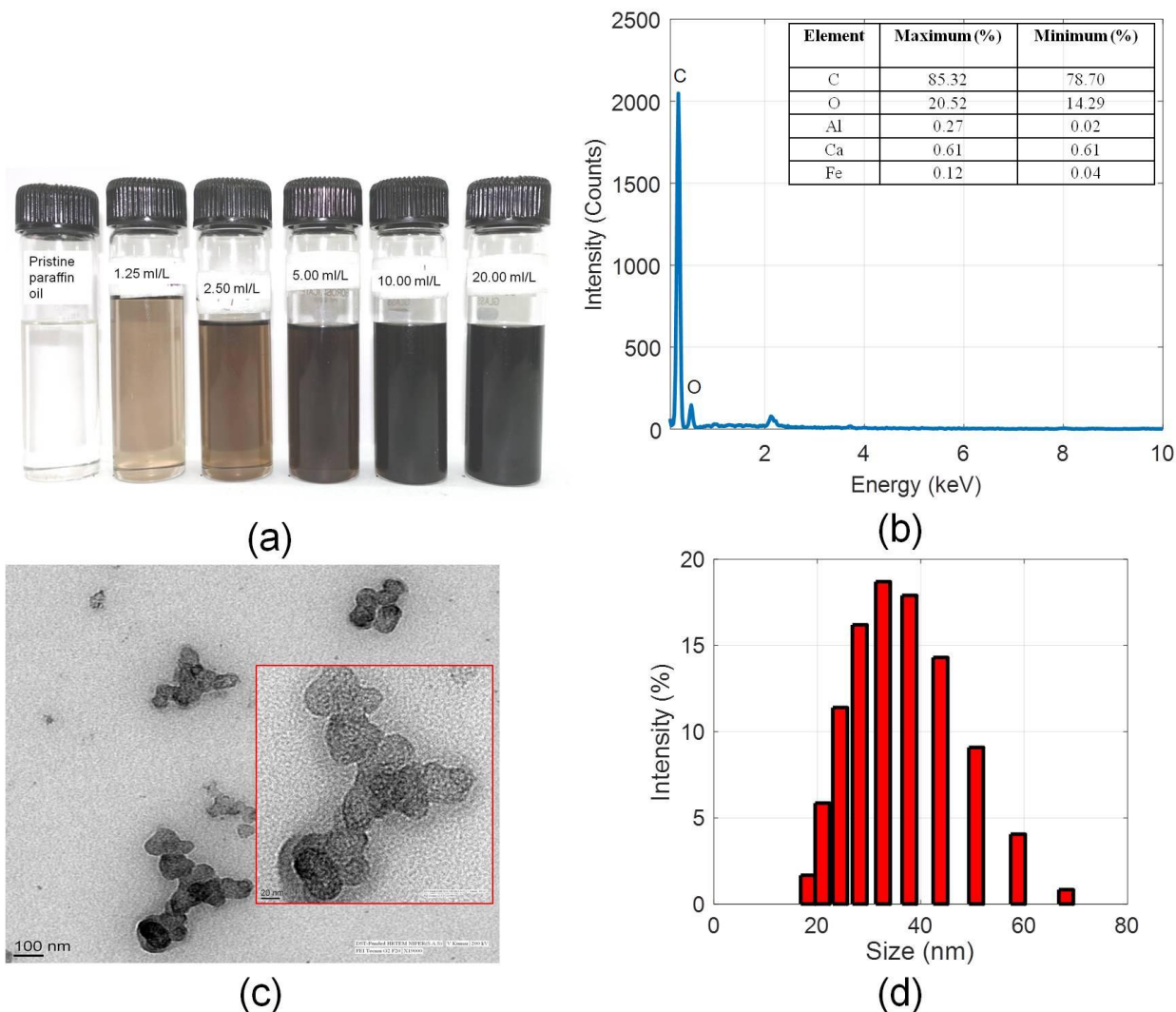

Fig. 2 (a) Photographs of the as-prepared nanofluids of different concentrations, (b) EDS of the residue left after evaporating 20 ml/liter nanofluid sample, (c) TEM images of the soot particles in the used engine oil, and (d) DLS measurements of the as-prepared nanofluid sample (5 ml/L).

In order to clearly gauge the photo-thermal conversion efficiency of the as-prepared nanofluids, laboratory scale experiments have been carefully designed. Nanofluids (housed in a cylindrical column with reflective surface at the bottom) of different concentrations were illuminated with a broad spectrum white light source [see Fig. 3(c)]. Samples were illuminated until these reached steady state temperatures. These measured steady state temperatures (averaged across the entire depth of the nanofluid column) essentially represent the photo-thermal conversion efficiencies of various nanofluid concentrations under optical heating. Figure 3(d) clearly points out that nanofluids have higher steady state temperatures (on the order of ~31 higher) relative to the case of pure paraffin oil. Interestingly, the highest concentration nanofluid (20 ml/L) does not have the highest photo-thermal conversion efficiency; instead it is highest for the nanofluid with a moderate concentration (2.5 ml/L). This could be understood from the spatial temperature distribution across the depth of the nanofluid column for various nanofluid concentrations (see Fig. 4). Spatial temperature field gives us insights into the photo-thermal conversion process. For a fixed physical thickness of the fluid layer, with increase in nanoparticle concentration, the photo-thermal conversion process tends to be limited to only top layers; not allowing the light to reach the lower layers-hence resulting in lower average steady state temperatures at very high concentrations.

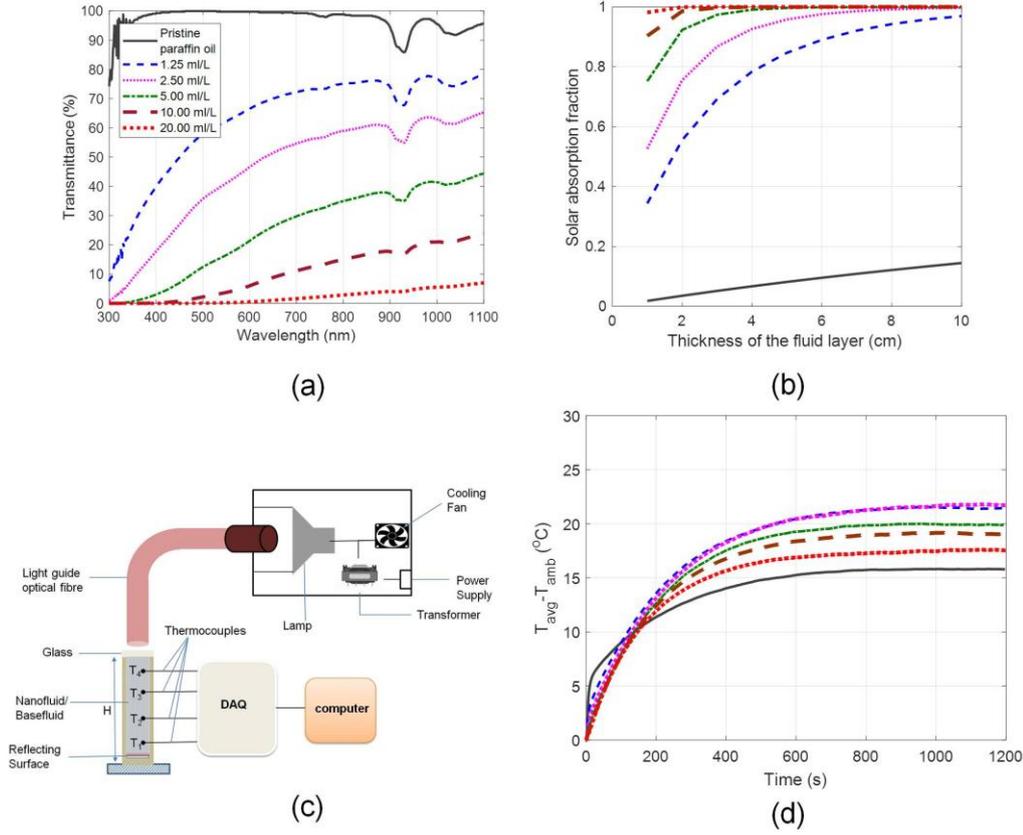

Fig. 3 (a) Spectral transmittance of the as-prepared nanofluids in the UV-VIS-NIR region, (b) Solar absorption fraction for various nanofluid concentrations as a function of fluid layer thickness, (c) schematic showing the experimental set-up for photo-thermal conversion experiments, and (d) steady state temperatures for various concentrations of nanofluids. $T_{avg} = (T_1 + T_2 + T_3 + T_4)/4$

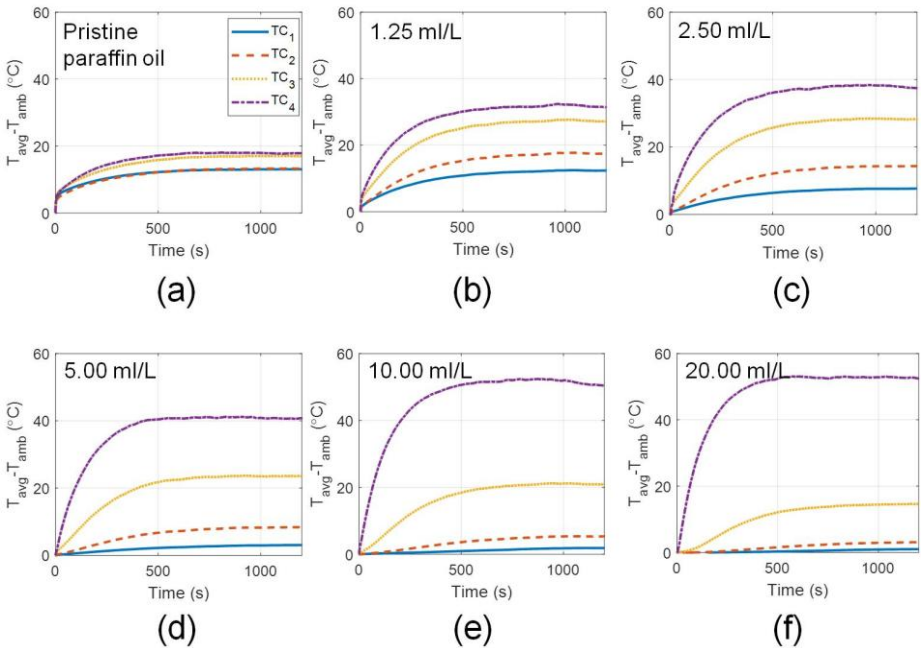

Fig. 4 Spatial temperature distribution, when the as-prepared nanofluids are illuminated under a white light source. (a) Pristine paraffin oil, (b) 1.25 ml/L, (c) 2.50 ml/L, (d) 5.00 ml/L, (e) 10.00 ml/L, and (f) 20.00 ml/L.

**Thermo-physical properties**

Thermal conductivity and viscosity are amongst the key thermo-physical properties that impact the redistribution of the absorbed energy within the fluid and pumping power requirements respectively. The as-prepared nanofluids show thermal conductivity enhancement [see Fig. 5(a)] of typically 2-4% (relative to pure paraffin oil); although not a significant enhancement, but could prove to be beneficial under high solar flux conditions. Viscosity measurements show linear increase in the viscosity with increase in concentration of the nanofluid, the increase being not greater than 2.5% [see Fig. 5(b)] even for the highest concentration (20 ml/L). Therefore, there shall not be requirement of additional pumping power when using these fluids in actual practice.

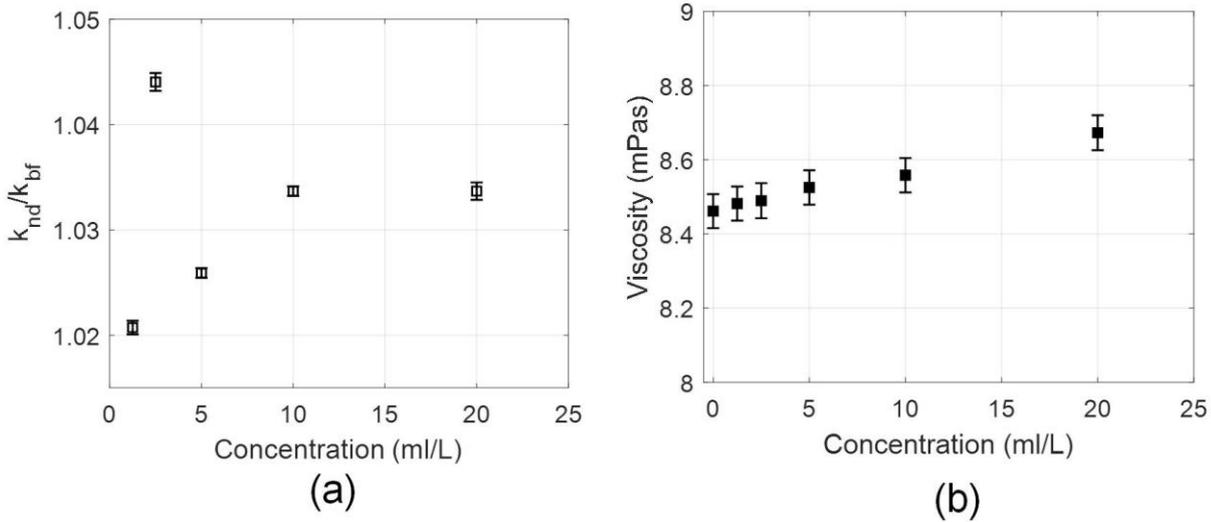

Fig. 5 (a) Thermal conductivity ratio ($k_{nd}/k_{bf}$), and (b) kinematic viscosity as a function of nanofluid concentration. Error bar represents the standard deviation.

**Stability of the as-prepared nanofluids**
*Long-term stability*

As 'heat transfer fluids' in volumetric absorption platforms; nanofluids are expected to maintain their optical and thermo-physical properties under extended periods without any appreciable degradation for consistent and efficient photo-thermal conversion. Long-term stability of the as-prepared nanofluid dispersions has been assessed under natural and accelerated sedimentation (centrifugation) conditions. Furthermore, during centrifugation, the fluid experiences severe shear stresses [35-40]- simulating real flow conditions which the heat transfer fluid may be subjected to in actual solar thermal systems. Figure 6 reveals that the as-prepared nanofluids exhibit remarkable long term stability and can withstand high shear stresses without any appreciable change in their optical characteristics, and nanoparticle hydrodynamic size distribution.

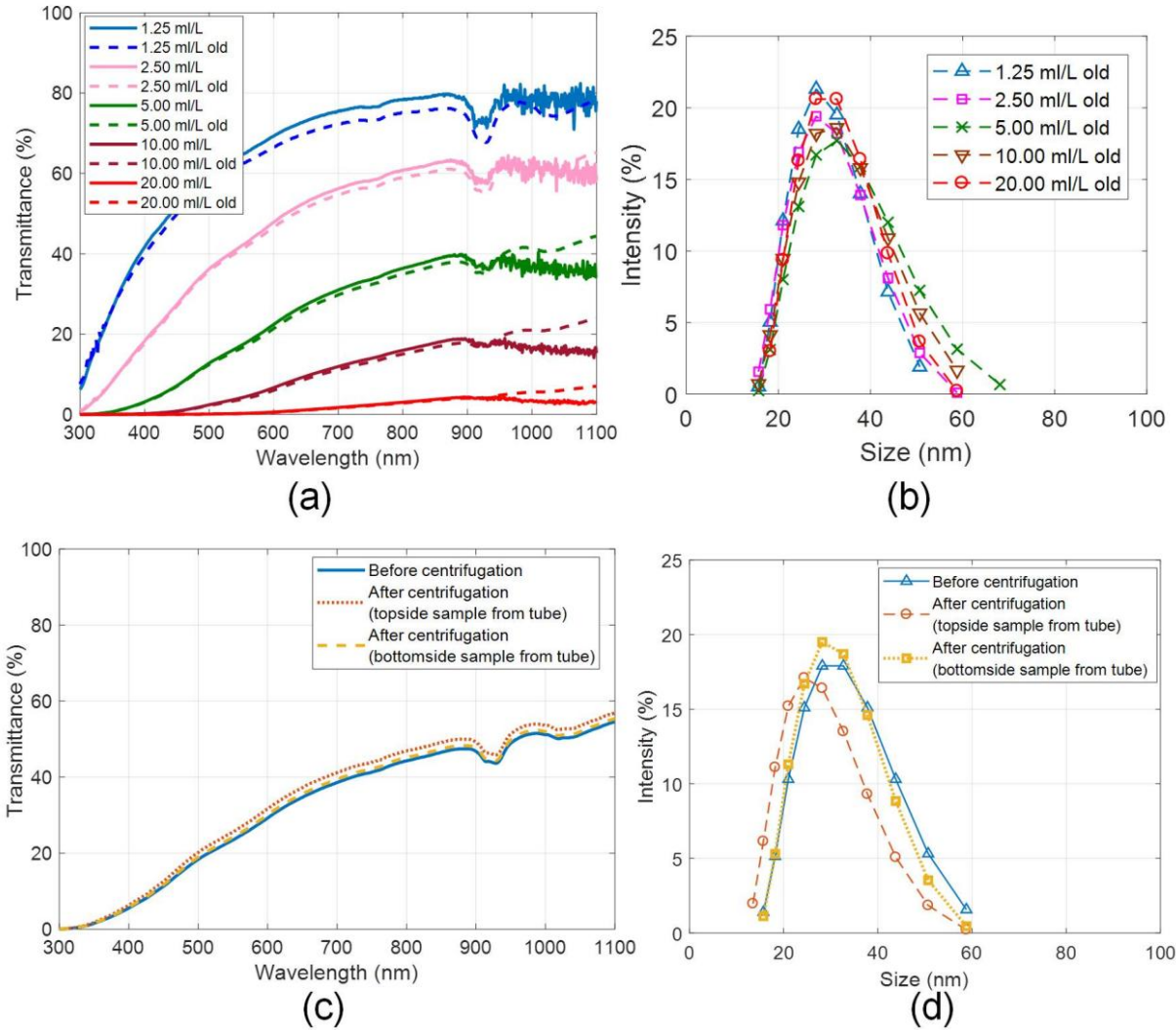

Fig. 6 (a) Spectral optical characteristics, and (b) hydrodynamic particle size distribution of the as-prepared and six months old nanofluid samples; and (c) spectral optical characteristics, and (d) hydrodynamic particle size distribution of the as-prepared and after centrifugation nanofluid samples (5ml/L).

*High temperature stability*
In addition to high photo-thermal conversion efficiency, the nanofluids should maintain their desired characteristics under constant and cyclic thermal loads. In its service life, the nanofluids are expected to absorb high solar flux (particularly in high solar concentration solar thermal systems), which in turn shall result in rapid and significant temperature rise. Furthermore, the nanofluid shall transfer the absorbed energy to a secondary fluid (such as water) - thus experiencing rapid temperature drops. These rapid heating and cooling cycles form the integral part of any power cycle in general and solar electricity generation systems (SEGS) in particular. The as-prepared nanofluids were found to posses excellent stability and retain their functional characteristics under constant as well as cyclic thermal loads. During cyclic thermal loading, the nanofluid was rapidly heated to a particular fixed temperature and then was suddenly cooled by dipping it into the water bath maintained at room temperature (see Fig. 7). For the purpose of tracking the temperatures, K-type thermocouple remained dipped in the nanofluid during the entire testing period. This however allowed the ambient air (oxygen) to interact with the nanofluid which in effect resulted in the oxidation of the basefluid (paraffin oil, see Fig. 8) - proving to be detrimental to the stability of the nanofluids particularly at high temperatures (see Figs. 9 and 10). Similar phenomenon was discovered during constant thermal loading (12 hour heating at constant temperature) prolonged heating as well (see Fig. 11) - indicating that it is not the thermal stresses but oxidation of the paraffin oil that renders the nanofluid unstable at high temperatures. Instructively, when the constant heating tests were carried out by making the container housing the nanofluid 'airtight'; no agglomeration or settling of the nanoparticles was observed [see Figs. 12(a) and (b)]. Moreover, nanoparticle size distribution and optical characteristics were retained even after constant heating for 12 hours at 300°C [see Figs. 12(c) and (d)].

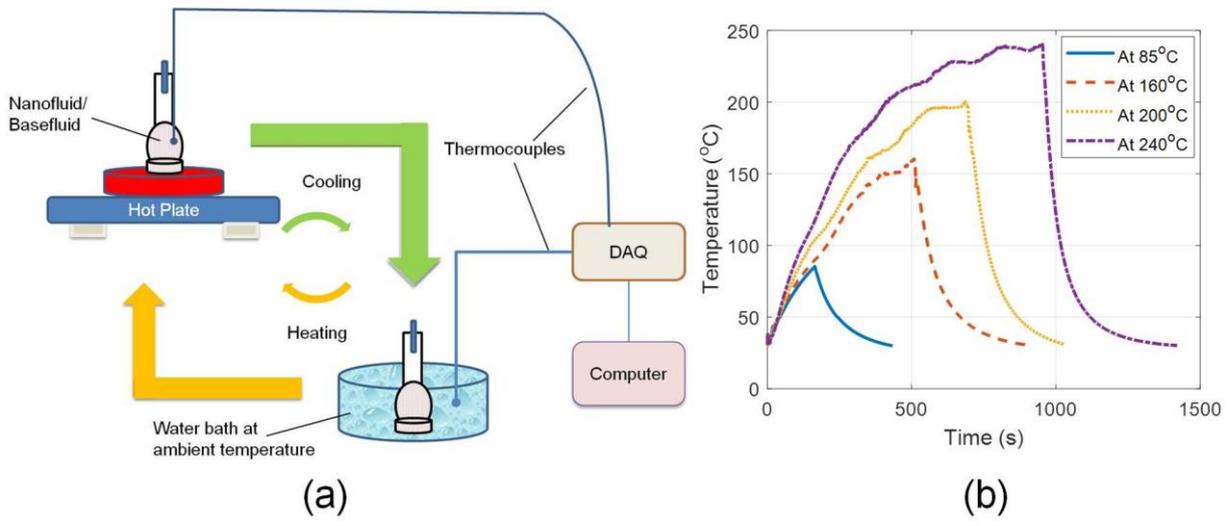

Fig. 7(a) Schematic showing the procedure followed for carrying out thermal cyclic tests, and (b) heating - cooling curves for the thermal cycling at different temperatures.

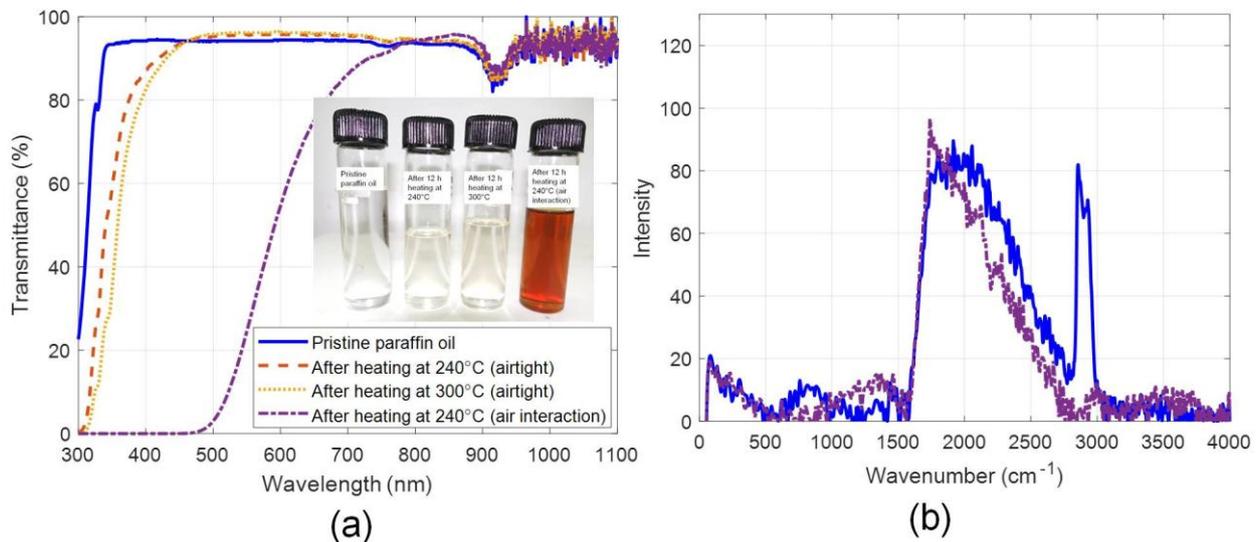

Fig. 8 (a) Effect of oxidation on the (a) optical properties, and (b) Raman spectra of pristine paraffin oil.

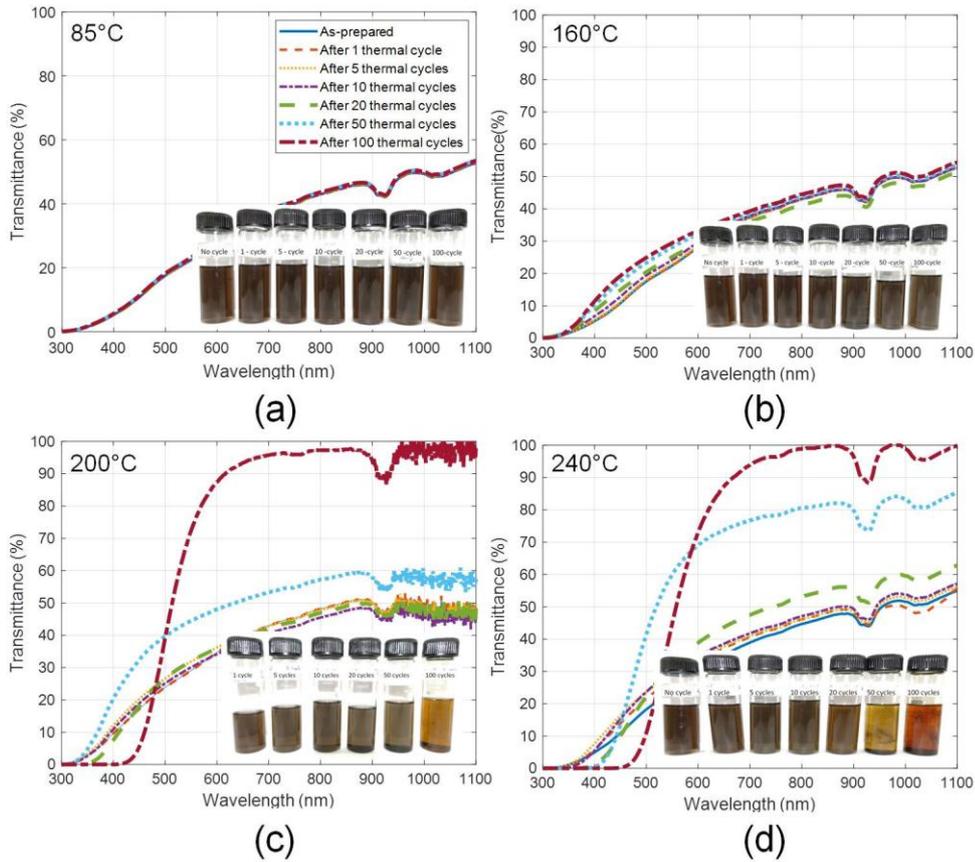

Fig. 9 (a) Effect of cyclic thermal loads on the optical properties of the as-prepared nanofluids at (a) 85°C, (b) 160°C, (c) 200°C, and 240°C.

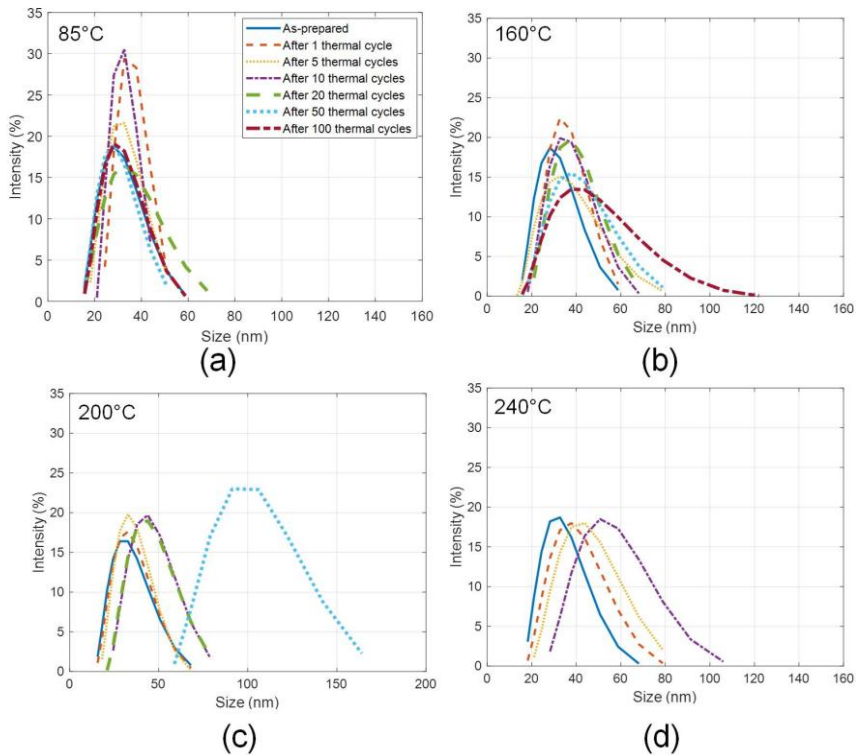

Fig. 10 (a) Effect of cyclic thermal loads on the hydrodynamic size distribution of the as-prepared nanofluids at (a) 85°C, (b) 160°C, (c) 200°C, and 240°C.

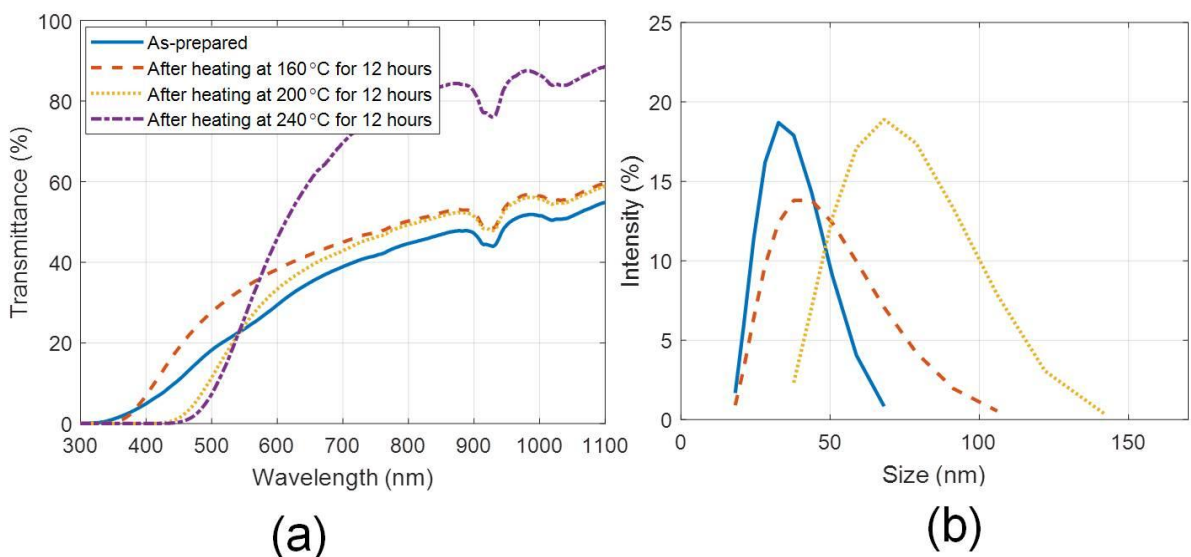

Fig. 11 (a) Effect of prolonged heating (for 12 hours) on the (a) optical properties, and (b) hydrodynamic size distribution of the as-prepared nanofluid (5 ml/L).

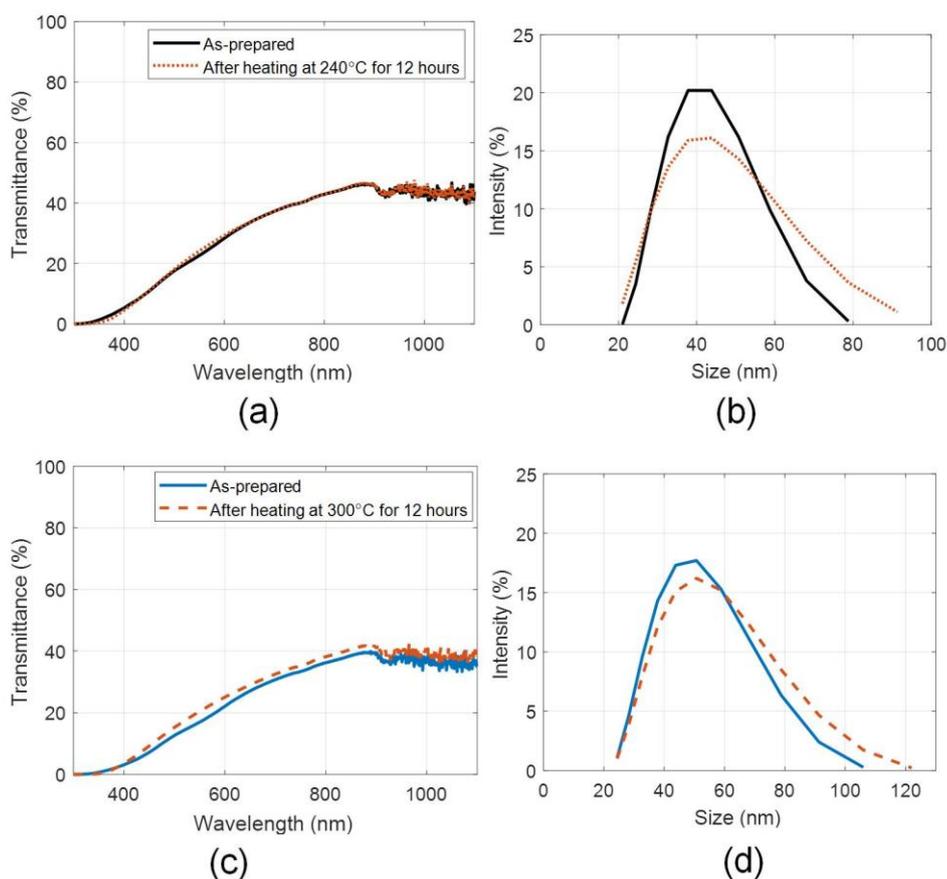

Fig. 12 (a) Spectral optical characteristics, and (b) hydrodynamic particle size distribution of the as-prepared and after heating (12 h at 240°C) nanofluid samples; and (c) spectral optical characteristics, and (d) hydrodynamic particle size distribution of the as-prepared and after heating ( for 12 hours at 300°C) nanofluid samples (5ml/L).

*Stability under ultra-violet light exposure*
Although ultraviolet radiations form only a small fraction of the incident solar energy; but given the fact that these are very high energy radiations, and may amount to significant values in case of concentrating solar thermal systems - the as-prepared nanofluids were tested exclusively under UV exposure. Interestingly, the nanofluids were found to be stable and retain their properties even after prolonged exposure to UV radiations (see Fig 13). This is a significant improvement, as exposure to UV radiations has been known to significantly impact the stability of the nanofluids, i.e., extensive agglomeration and settling of the nanoparticles occurs when exposed to UV radiations [28].

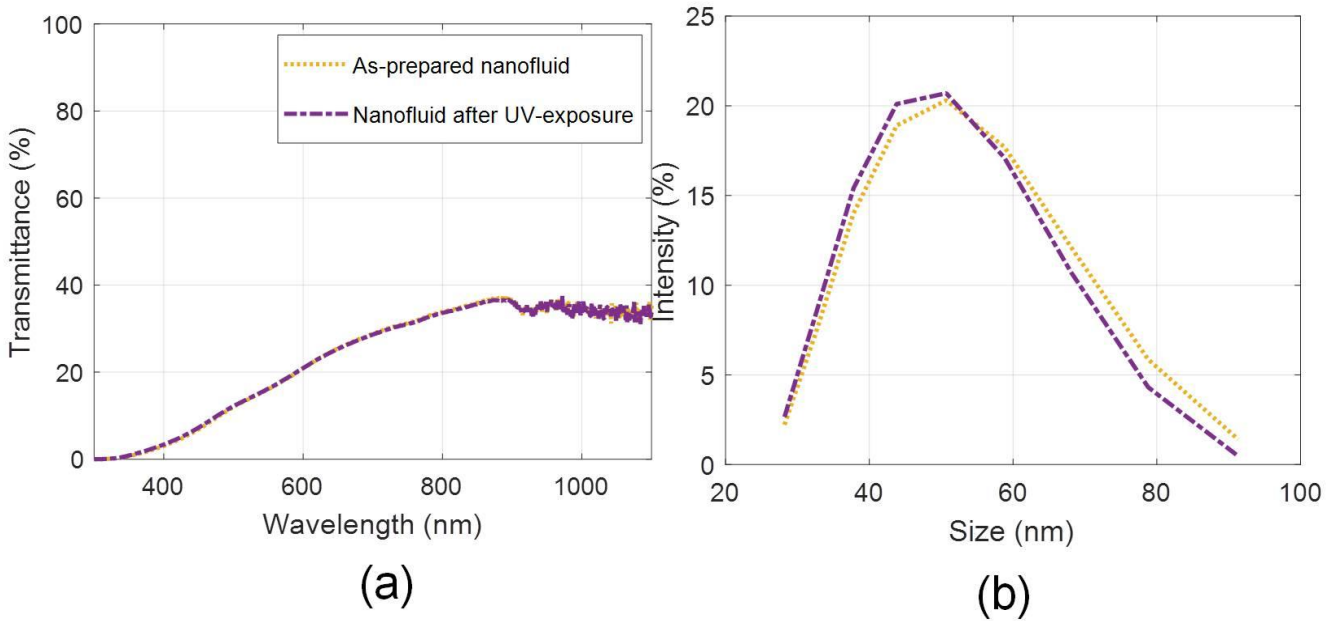

Fig. 13 (a) Effect of UV exposure (6 hours) on the (a) optical properties, and (b) hydrodynamic particle size distribution of the as-prepared nanofluid (5ml/L).

**Design and performance of the as-prepared nanofluid based volumetric absorption solar thermal platform**
Rigorous testing of the as-prepared nanofluids has revealed that indeed these could operate under real world service conditions without losing their functional properties. Building upon this, we have carefully designed a volumetrically absorbing solar thermal system, wherein a linear Fresnel lens is used to concentrate the incident normal solar irradiance onto the receiver lying along the focal line of the concentrator. The receiver is essential a closed rectangular conduit in which the fluid is made to flow. The conduit has been so designed that it could be employed both in surface as well as volumetric absorption modes - the internal three sides (bottom and sides) being 'state of the art' solar selective surfaces (having high solar weighted absorptivity, ~0.96; and low infrared emissivity, ~0.12) and the top side being made of glass to allow the sunlight to pass through and reduce thermal losses. Furthermore, the outside three sides have been insulated to reduce thermal losses [see Fig. 14(a)]

In surface absorption mode, pure paraffin oil was made to flow in the receiver. Paraffin oil being nearly transparent in the solar wavelength region, allows the sunlight to interact with the solar selective surfaces. The absorbed solar energy is then transferred to the fluid through convection and conduction mechanisms. This configuration essentially simulates the heat transfer mechanisms involved in typical incumbent solar thermal systems.

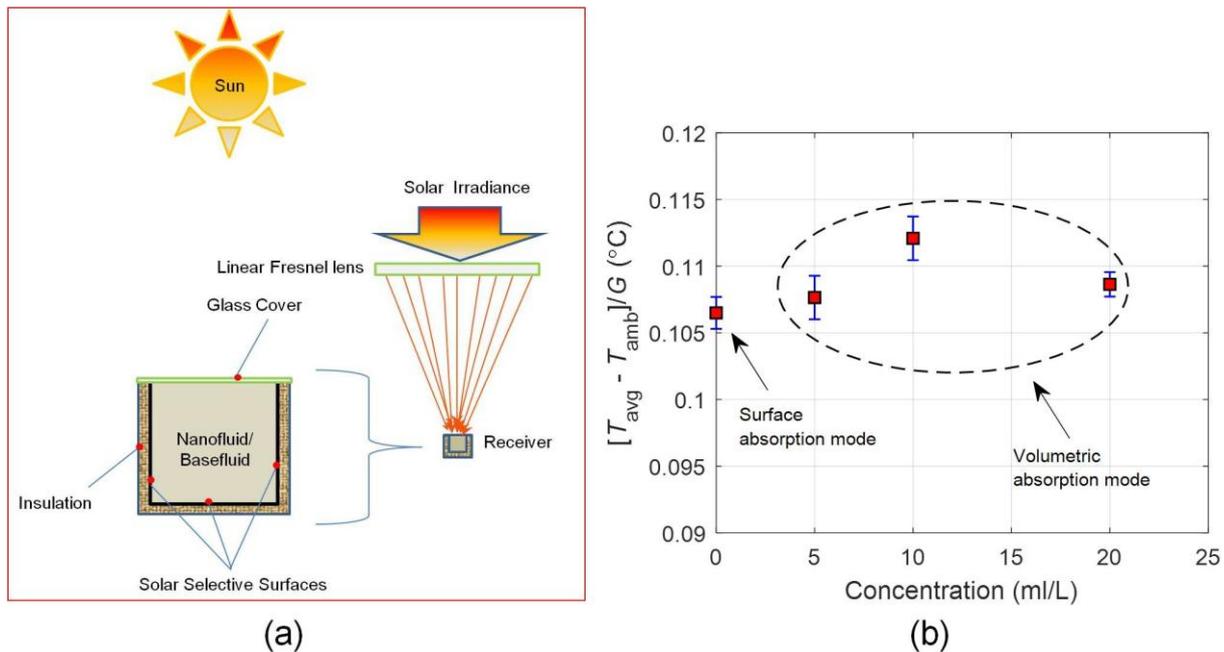

Fig. 14 (a) Schematic showing the design of the volumetric absorption platform, and (b) Steady state temperatures under surface and volumetric (at various nanofluid concentrations) absorption modes.

On the other hand, in volumetric absorption mode, the as-prepared nanofluids (of different concentrations) were made to flow through the receiver. Herein, the nanofluid directly absorbs the solar energy - nanoparticles owing to their broad wavelength absorption characteristics; convert the solar energy into thermal energy through non-radiative decay of the absorbed photons. Subsequently, the absorbed energy is transferred to the surrounding medium at staggeringly rapid rate (owing to the extremely small size of the particles). This results in efficient photo-thermal conversion of the incident solar energy.

Comparison between the two absorption modes (surface and volumetric) clearly points out that indeed higher steady state temperatures (~5% higher) could be achieved in case the sunlight is allowed to directly interact with the nanofluid i.e., volumetric absorption outperforms the surface absorption case under real world flow conditions and under the sun [see Fig. 14(b)] Furthermore, careful observation of the graphs tell us that it is imperative to choose the right concentration of the nanofluids (for a given fluid layer thickness) in order to maximize the resulting steady state temperatures and hence the performance characteristics.

As a whole, the present work reports a simple, cost effective and scalable method to synthesize broad absorption nanofluids from 'used engine oil'. These as-prepared nanofluids have shown to possess remarkable properties in relation to its candidature as a potential working fluid in volumetric absorption solar thermal systems. Building on it, an efficient solar thermal platform has been devised which gives higher steady state temperatures in volumetric absorption mode. Thus we have been able to engineer efficient volumetric absorption solar thermal platforms which outperform their surface absorption counterparts (conventional solar thermal systems) under realistic conditions and under the sun.

## METHODS

**Nanofluid preparation.** Used engine oil has been collected from a 15000 km run 4-stroke diesel engine. To filter out sludge, resin etc. a cotton cloth has been used. Subsequently, the filtered 'used engine oil' was furthered filtered with 0.7um filter paper. Desired fractions of the resulting filtered used oil were then mixed into pure paraffin oil followed by 30 minutes of ultra-sonication in a bath type ultasonicator - thus forming nanofluids of different concentration.

**Characterization and measurements**
EDS. Used engine oil cannot be directly analyzed by EDS because of the presence of oil which may contaminate the electron beam. So sample was prepared by evaporating 20ml/liter sample at 160°C and the leftover after evaporation was collected. The collected solid particles were then washed with the ethanol (5 times) in order to remove any traces of oil. Finally, the washed particles were loaded onto the copper grid for EDS analysis.

Spectral analysis. Transmittance measurements in the UV-VIS-NIR region were done using spectrophotometer. Shimazdu UV-2600

TEM. The sample for TEM analysis was prepared by solvent extraction method. The used oil was mixed with n-heptane (1:60) and ultra-sonicated for 30 minute. The prepared sample was then placed on the carbon grid and washed with n-heptane to remove any traces of oil on the surface of soot particles. Furthermore, to get the better image quality, sample was washed with diethyl ether (2 times). TEM, FEI Tecnai G2 F20, Netherlands

DLS. Hydrodynamic size distribution measurements were made using Malvern Zetasizer Nano S (ZEN 1600)

Thermal conductivity of the as-prepared nanofluid was measured by KD2 pro which works on the transient hot wire method.

Raman Spectra. Measured at 532 nm, Horiba Scientific

Viscosity measurements were made using capillary action viscometer.

White light source. Light guide connected to a 3200K Color temperature halogen lamp (250 W), Philips

UV light source. 125 W lamp, Philips

Temperature measurements. K type thermocouples and infrared camera, Keysight

Data Acquisition (DAQ). DAQ card NI 9123, Chassis NI 9721

Incident solar flux. Ring shaded Pyranometer, Kipp & Zonen

Heating sources. Metal top and ceramic based hot plates

Ultra-sonicator. Bath type - 250 W, Sarthak Scientific

**Constructional and operational parameters of the volumetric absorption solar thermal platform**
Linear Fresnel lens. 2 in number, each of length 500 mm, width 500 mm PMMA-3t, NTKJ, Japan
Solar selective surface. Black chrome coated copper sheet, Solchrome
Dimensions of the receiver (internal). length 1000 mm, width 29 mm, and height 22 mm
Volume flow rate. 0.5 lpm

**Acknowledgements**
This work is supported by Department of Science and Technology - Science and Engineering Research Board (Sanction order no. ECR/ 2016/000462). Authors also acknowledge the support provided by Mechanical Engineering Department and the Department of Biotechnology at Thapar Institute of Engineering and Technology. Support provided by the Mechanical Engineering Department at IIT Ropar is also gratefully acknowledged.